\documentclass[prd,preprintnumbers,amsmath,amssymb,usenatbib,nofootinbib,superscriptaddress,showkeys,showpacs,11pt]{revtex4-2}

\usepackage{amsmath}
\usepackage{amsfonts,color}
\usepackage{amssymb,float}
\usepackage{graphicx}
\usepackage{subfigure}
\usepackage{appendix}
\usepackage[colorlinks]{hyperref}
\usepackage{xcolor}
\usepackage{orcidlink}
\usepackage{epsf}
\usepackage{bm}
\usepackage{epstopdf}
\usepackage{natbib}
\usepackage{lipsum}

\begin{document}

\title{Observational Hubble parameter data constraints on the interactive model of $f(T)$ gravity with particle creation}

\author{S. Ganjizadeh}
\email{st.s.ganjizadeh@iauamol.ac.ir}
\affiliation{Department of Physics, Ayatollah Amoli Branch, Islamic Azad University, Amol, Iran}

\author{Alireza Amani\orcidlink{0000-0002-1296-614X}}
\email{al.amani@iau.ac.ir}
\affiliation{Department of Physics, Ayatollah Amoli Branch, Islamic Azad University, Amol, Iran}

\author{M. A. Ramzanpour}
\email{m.ramzanpour@iauamol.ac.ir}
\affiliation{Department of Physics, Ayatollah Amoli Branch, Islamic Azad University, Amol, Iran}

\date{\today}

\begin{abstract}

In this paper, we consider an open system from the thermodynamic perspective for an adiabatic FRW universe model in which particle creation occurs within the system. In that case, the modified continuity equation is obtained and then we correspond it to the continuity equation of $f(T)$ gravity. So, we take $f(T)$ gravity with the viscous fluid in flat-FRW metric, in which $T$ is the torsion scalar. We consider the contents of the universe to be dark matter and dark energy and consider an interaction term between them. The interesting point of this study is that we make equivalent the modified continuity equation resulting from the particle creation with the matter continuity equation resulting from $f(T)$ gravity. The result of this evaluation creates a relationship between the number of particles and the scale factor. In what follows, we write the corresponding cosmological parameters in terms of the number of particles and also reconstruct the number of particles in terms of the redshift parameter, then We parameterize the Hubble parameter derived from power-law cosmology with 51 data from the Hubble observational parameter. Next, we plot the corresponding cosmological parameters for the dark energy in terms of the redshift to investigate the accelerated expansion of the universe. In addition, by using the sound speed parameter, we discuss the stability analysis and instability analysis of the present model in different eras of the universe. Finally, we plot the density parameter values for dark energy and dark matter in terms of the redshift parameter.
\end{abstract}



\maketitle
\newpage
\section{Introduction}\label{I}

In the late twentieth century, by measuring cosmic distances and observing cosmic objects beyond the Milky Way, we realized that the universe is rapidly expanding, and this event occurs when there is a force greater than gravity but in the opposite direction in the universe. But the origin of this mysterious force is introduced by an unknown energy called dark energy, which is supposed to surround the entire space and increase the expansion rate of the universe. In fact, this energy was first discovered in type-Ia supernovae and then reconfirmed by two approaches to the cosmic microwave background radiation and the large-scale structure of the universe \cite{Riess_1998, Perlmutter_1999, Bennett_2003, Tegmark_2004}. The problem of dark energy originates from a fundamental theoretical framework such as string theory and quantum gravity. But despite the evidence that the universe is accelerating, these observations do not fit the standard model of cosmology within framework of general relativity. To resolve this contradiction, we introduce a factor called negative pressure for the dark energy source, which can describe the observed accelerated expansion of the universe. Extensive studies of dark energy have been explored with models such as cosmological constant, scalar fields, modified gravity, agegraphic, holography, and teleparallel gravity  \cite{Weinberg-1989, Caldwell-2002, Amani-2011, Sadeghi1-2009, Setare-2009, Setare1-2009, Battye-2016, Li-2012, Pourhassan-2014, Amani1-2015, Faraoni-2016, Sadeghi1-2016, Wei-2009, Amani-2015, Nojiri_2007, Li-2004, Campo-2011, Hu-2015, Amani-2013, Amani-2014, Naji-2014, Morais-2017, Zhang1-2017, SadeghiKhurshudyanJ1-2014, Sadeghi-2010, Sadeghi-2009, Amani-2016, Singh-2016, Sahni1-2003, Setare-2008, Brito-2015, Setaremr-2008, Amanifarahani-2012, Amanifarahani1-2012, Amanipourhassan-2012,  Bhoyar-2017, Chirde-2018, Singh-2018, Nagpal-2018, Amani1-2011, Wei-2008, Wei-2007}.

In addition, another scenario called teleparallel gravity, first proposed by Einstein, was introduced for the unification between electromagnetism and gravity. In this theory, space-time is characterized by a linear connection without curvature by a metric tensor field called the dynamical tetrad field. Therefore, the geometry of the teleparallel theory uses the linear Weitzenb\"{o}ck connection, to represent a torsion connection without curvature. But the geometry of general relativity is based on the Levi-Civita connection to represent a curvature connection without torsion \cite{Einstein_1928, Weitzenbock_1923, Linder_2010, Myrzakulov_2011, Li_2011, Myrzakulov_2012, Harko_2011, Nojiri_2005}.

As mentioned above, current observations with different scenarios suggest that the universe is undergoing accelerated expansion, which leads to the crossing of the phantom divided-line (equation of state (EoS) is less than -1). Furthermore, the particle creation mechanism is an appropriate alternative to describe the current acceleration of the universe, so that it can successfully mimic the standard cosmology. In that case, from the point of view of the first law of thermodynamics, the creation of a dynamical number of particles, $N$, leads to a modified continuity equation for a system called the adiabatic universe \cite{ZelDovich-1972, Prigogine-1989, Parker-2012, Cardenas-2012, Fabris-2014, Biswas-2017, Saha-2017, Tu-2020, Mandal-2020}. Note that the corresponding continuity equation is also obtained via Einstein's equation in standard cosmology as $\dot \rho + 3 H (\rho + p) = 0$, in which $\rho$, $p$, and $H$ are energy density, pressure, and Hubble parameter, respectively. An important point that should be noted in the mechanism of particle creation is that its continuity equation, in addition to standard terms, also has the term of the rate of the number of particles as ($\frac {\dot N}{N} (\rho+p)$). A number of papers attribute the origin of particle number rate to the bulk viscosity. In the event that, an open thermodynamic system in which the number of fluid particles is not conserved leads to a modified particle conservation equation, known as the equilibrium equation for particle flow. Hence, we will consider that the corresponding particle flow is related to the energy flow between the dark components of the universe. Then, the particle flux resulting from particle creation depends on the energy transfer between dark energy and dark matter. In that case, we consider this issue as energy transfer, so-called interaction between the dark components of the universe.

It should be noted that the particle creation in the black hole and the expanding universe was also studied \cite{Hawking_1975, Parker-1968}. For this purpose, in this work, we intend to implement the mechanism of the particle creation to describe the accelerated expansion of the universe. This motivated us to make a correspondence between the particle creation mechanism and the $f(T)$ gravity model. Therefore, using this correspondence and a power-law model for the scale factor, we expect to examine the accelerated expansion of the universe with observational data.

This paper is organized as follows:

In Sec. \ref{II}, we implement the fundamental formation of viscous $f(T)$ gravity in the flat-FRW metric. In Sec. \ref{III}, we consider that the universe is dominated by a bulk viscosity fluid, and also obtain the Friedmann equations and EoS in an interacting model. In Sec. \ref{IV}, we first explain the mechanism of particle creation, and then using the interaction term, we correspond the particle creation with the viscous $f(T)$ gravity model. In what follows, in Sec. \ref{V}, we investigate the present model by observational constraints with Hubble data, and then the cosmological parameters will plot in terms of the redshift parameter. In Sec. \ref{VI}, we explore the stability conditions for the corresponding model. Finally, in Sec. \ref{VII}, we will present a conclusion for the present model.


\section{Foundation of $f(T)$ gravity}\label{II}

In this section, we are going to review teleparallel gravity. For this purpose, we can write the corresponding action for $f(T)$ gravity in the following form
\begin{equation}\label{action1}
S = \int e \left(\frac{f(T)}{2 \kappa^2} + \mathcal{L}_m \right) d^4x,
\end{equation}
where $\kappa^2=8 \pi G$, and $e^i_{\,\,\mu}$ is vierbein field from which obtains determinant of vierbein field as $e=det\left( e^i_{\,\,\mu}\right)$, and $\mathcal{L}_m$ is the matter Lagrangian density. For a more complete explanation, we write the basis of vierbein field $e_i(x^\mu)$ as the tangent vectors at point $x^\mu$ of a manifold as $e_i . e_j = \eta_{ij}$, in which we consider the corresponding signature of Minkowski space-time as $\eta_{ij}=diag(+1,-1,-1,-1)$. In that case, by using of the relationship $e_i = e_i\,^\mu \partial_\mu$, we can obtain metric tensor $g_{\mu \nu}$ by the tetrad or vierbein field $e^i_{\,\,\mu}$ as $g_{\mu \nu}=\eta_{\mu \nu} e^i_{\,\,\mu} e^i_{\,\,\nu}$, which in turn leads to be the relationships $e_i\,^\mu e^i\,_\nu=\delta_\nu^\mu$ and $e_i\,^\mu e^j\,_\mu=\delta_i^j$. It should be noted that the Greek and Latin letters denote components of space-time coordinates and  components of tangent space-time coordinates, respectively.

As we know, we use the Weitzenb\"ock connection in teleparallel gravity instead of the Levi-Civita connection in general relativity as
\begin{equation}\label{kris1}
 \Gamma^\lambda_{~\mu\nu} = e_i^{~\lambda} \partial_\mu e^i_{~\nu} = -e^i_{~\mu} \partial_\nu e_i^{~\lambda}.
\end{equation}

By using this connection, we can introduce the torsion tensor and the torsion scalar in the following form
\begin{subequations}\label{TTmn1}
\begin{eqnarray}\label{torTS1}
&{T^\lambda }_{\mu \nu } = \Gamma^\lambda_{~\mu\nu} - \Gamma^\lambda_{~\nu \mu} =  {e_a}^\lambda \,\left( {{\partial _\mu }{e^a}_\nu  - {\partial _\nu }{e^a}_\mu } \right), \label{torTS1-1}\\
&T = {S_\lambda }^{\mu \nu }\,{T^\lambda }_{\mu \nu }, \label{torTS1-2}
 \end{eqnarray}
\end{subequations}
where
\begin{subequations}\label{SK1}
\begin{eqnarray}
 &S_\lambda ^{\mu \nu } = \frac{1}{2}\,\left( {K_\lambda ^{\mu \nu } + \delta _\lambda ^\mu \,T_\alpha ^{\alpha \nu } - \delta _\lambda ^\nu \,T_\alpha ^{\alpha \mu }} \right), \label{SK1-1}\\
 &K_\lambda ^{\mu \nu } =  - \frac{1}{2}\,\left( {{T^{\mu \nu }}_\lambda  - T_\lambda ^{\nu \mu } - T_\lambda ^{\mu \nu }} \right), \label{SK1-2}
 \end{eqnarray}
\end{subequations}
to be known as the antisymmetric tensor and the contortion tensor, respectively. Also, we can obtain the Ricci tensor and the Ricci scalar by aforesaid relations in the following form
\begin{subequations}\label{Riccits1}
\begin{eqnarray}
 & R_{\mu \nu} = -\nabla^\rho S_{\nu \rho \mu} - g_{\mu \nu} \nabla^\rho T^\sigma_{~ \rho \sigma} - S^{\rho \sigma}_{~ ~ \mu} K_{\sigma \rho \nu}, \label{Riccits1-1}\\
 &R = -T - 2 \nabla^\mu \left(T^\nu_{~ \mu \nu}\right). \label{Riccits1-2}
 \end{eqnarray}
\end{subequations}

By taking variation of the action \eqref{action1} with respect to the tetrad field, the field equations obtain as
\begin{eqnarray}\label{eom_fT_general}
e^{-1} \partial_{\mu}\left(e\, e^\lambda_a\,
S_{\lambda}^{~\mu\nu}\right) f_{T} - e_{a}^{\rho} \, T^{\lambda}_
{~\mu\rho} S_{\lambda}{}^{\nu\mu} f_{T}
 + e^\lambda_a S_{\lambda}{}^{\mu\nu}\, \partial_{\mu}(T) f_{TT} +\frac{1}{4}e_{a}^{\nu} f
 = \frac{1}{2} \kappa^2 e_{a}^{\lambda} \mathcal{T}_{\lambda}{}^{\nu},
\end{eqnarray}
where index $T$ represents the derivative with respect to torsion scalar, and $\mathcal{T}_{\lambda}{}^{\nu}$ is the energy-momentum tensor.
Here we consider the background manifold as flat Friedmann-Robertson-Walker (FRW) metric as
\begin{equation}\label{metricfrw1}
 ds^{2} = dt^{2} - a^{2}(t) \left(dx^2 +dy^2 + dz^2 \right),
\end{equation}
where $a(t)$ is scale factor. Then, one can immediately obtain the tetrad field as $e_{\mu}^a=\mathrm{diag}(1,a,a,a)$ or $e^{\mu}_a=\mathrm{diag}(1,a^{-1},a^{-1},a^{-1})$, and the torsion scalar obtains in the following form
\begin{eqnarray}\label{eamu1}
T = -6 H^2,
\end{eqnarray}
where $H = \frac{\dot{a}}{a}$ is introduced the Hubble parameter.


\section{Viscous $f(T)$ gravity}\label{III}

In this section, we intend to consider the current universe is dominated by an anisotropic fluid with the view that the universe is in a more realistic form. This means that we would like to describe the corresponding universe by $f(T)$ gravity in the presence of a bulk viscous fluid. Considering that the fluid viscosity indicates its resistance to flow as a dissipative phenomenon, so the pressure within the universe becomes negative due to it. Then, the presence of bulk viscosity causes the accelerated expansion of the universe \cite{Tripathy-2010}. In this case, we can write the energy-momentum tensor in the presence of bulk viscosity as follows:
\begin{equation}\label{Tij1}
\mathcal{T}_i^j=(\rho_{tot} + p_{tot} + p_b) u_i u^j - \left(p_{tot} + p_b\right)\,  \delta_i^j,
\end{equation}
where $\rho_{tot}$, $p_{tot}$, and $p_b$ are respectively introduced as the total energy density, the total pressure, and the pressure of bulk viscosity of fluid within the universe. We use the $4$-velocity $u_\mu$ notation as $u^i$ = (+1,0,0,0) which gives rise to $u_i u^j$ = 1. Therefore, we can acquire the non-zero energy-momentum tensor elements as
\begin{equation}\label{tau1}
 \mathcal{T}_i^j =diag \left( \rho_{tot}, - p_{tot} - p_b, - p_{tot} - p_b, - p_{tot} - p_b\right),
\end{equation}
where to substitution the aforesaid energy-momentum tensor elements and FRW metric into Eq.  \eqref{eom_fT_general}, the Friedmann equations obtain as
\begin{subequations}\label{fried1}
\begin{eqnarray}
 & \kappa^2 \rho_{tot} = 6 H^2 \partial_T f + \frac{1}{2} f,\label{fried1-1}\\
 & \kappa^2 \left({p}_{tot} +p_b\right) = -2 \left(\dot{H} + 3 H^2 \right) \partial_T f + 24 H^2 \dot{H} \partial_{T T} f - \frac{1}{2} f, \label{fried1-2}
\end{eqnarray}
\end{subequations}
where the dot and index $T$ indicate the derivative with respect to time evolution and torsion scalar, respectively. Note that the Friedmann equations \eqref{fried1} are written in the standard form in general relativity by applying condition $f(T) = T$ in a perfect fluid ($p_b = 0$). Therefore, we consider that the components of the universe include matter and dark energy in an anisotropic fluid ($p_b \neq 0$). In that case, we write down the total energy density and the total pressure in terms of the components of the universe in the following form
\begin{subequations}\label{fried2}
\begin{eqnarray}
& \rho_{tot} = \rho + \rho_{de},\label{fried2-1}\\
& {p}_{tot} = {p} + p_{de},\label{fried2-2}
\end{eqnarray}
\end{subequations}
where $\rho$ and $p$ are the energy density and the pressure of matter, and $\rho_{de}$ and $p_{de}$ are the energy density and the pressure of dark energy. However, from the Eqs. \eqref{fried1} and \eqref{fried2}, we obtain relations $\rho_{de}$ and $p_{de}$ in the following form
\begin{subequations}\label{fried3}
\begin{eqnarray}
& \kappa^2 \rho_{de} = 6 H^2 \partial_T f + \frac{1}{2} f - \kappa^2 \rho,\label{fried3-1}\\
& \kappa^2 {p}_{de} = -2 \left(\dot{H} + 3 H^2 \right) \partial_T f + 24 H^2 \dot{H} \partial_{T T} f - \frac{1}{2} f - \kappa^2 \rho \, \omega_m - 3 \kappa^2 \xi H, \label{fried3-2}
\end{eqnarray}
\end{subequations}
where $\omega_m = p/\rho$ is as the matter Equation of State (EoS) which is a constant, and term $3 \xi H$ is substituted with  the pressure of bulk viscosity, $p_b$, in which the coefficient $\xi$ is a positive constant. Afterward, we can introduce the EoS of dark energy as follows:
\begin{equation}\label{eos1}
\omega_{de} = \frac{p_{de}}{\rho_{de}} = -1 - \frac{2 \dot{H} \partial_T f - 24 H^2 \dot{H} \partial_{T T} f + \kappa^2 \rho \, (1+ \omega_m) + 3 \kappa^2 \xi H}{6 H^2 \partial_T f + \frac{1}{2} f - \kappa^2 \rho},
\end{equation}
where the corresponding EoS of dark energy describes the dynamic of the universe in late time, and depends on modified gravity, the matter, and viscous fluid.

In what follows, we can obtain the total continuity equation for the total fluid inside the universe as
\begin{equation}\label{continuity1}
\dot{\rho}_{tot}+3 H \left(\rho_{tot} + {p}_{tot} + p_b\right)=0,
\end{equation}
where one can immediately be written separately in terms of the universe components in the presence of an interacting term in the form
\begin{subequations}\label{continuity2}
\begin{eqnarray}
 &\dot{\rho}+3 H \left(\rho + p \right)= \mathcal{Q},\label{continuity2-1}\\
 &\dot{\rho}_{de}+3 H \left(\rho_{de} + {p}_{de} + p_b\right)= -\mathcal{Q},\label{continuity2-2}
\end{eqnarray}
\end{subequations}
where $\mathcal{Q}$ is introduced as an interaction term. Note that the interaction term appears when the energy flow is transferred between the dark components of the universe. It is obvious that the quantity $\mathcal{Q}$ should be positive, this is, the energy transfer from dark energy to dark matter occurs. Therefore, the positivity of the interaction term guarantees that the second law of thermodynamics is realized \cite{Pavon_2009}. In that case, since the unit $\mathcal{Q}$ is the inverse of the time evolution, it is natural choose this value as the product of the Hubble parameter and the energy density. Herein we consider $\mathcal{Q} = 3 b^2 H \rho$ in which $b^2$ is the coupling parameter or transmission intensity.

In the next section, we will discuss the thermodynamic study of the particle creation in an open system. And then, the corresponding open system is considered an adiabatic universe, so that, we will see, the particle creation is related to the interaction term between the dark parts of the universe.

\section{Correspondence between matter creation  and viscous $f(T)$ gravity}\label{IV}

As we know, the first law of thermodynamics or in other words the Gibbs equation with the dynamic number of particles N and the enthalpy density $h = \rho + p$ in an open system is written as follows:
\begin{equation}\label{fther1}
dE = dQ - p dV + \frac {h}{n} d(nV),
\end{equation}
where $\rho = \frac{E}{V}$ and $n = \frac{N}{V}$ are the energy density and the density of particles, respectively, where $E$ and $V$ are the internal energy and the volume of the system, respectively.

Now we consider the open system as an adiabatic FRW universe model as $dQ = 0$ with the existence of the creation or elimination of the number of particles in the corresponding system. In order to realize the creation of the particle, we consider the volume of the universe as a sphere with the radius of the scale factor $a(t)$ in the form of the relationship $V = \frac{4 \pi}{3}a^3$, so that, Eq. \eqref{fther1} yields
\begin{equation}\label{fther2}
\dot \rho - (\rho + p) \frac{\dot{n}}{n} = 0.
\end{equation}

By taking the derivative from the relationship $N = n V$ with respect to time evolution, we can clearly obtain the following relation
\begin{equation}\label{fther3}
\frac {\dot n}{n} = \frac {\dot N}{N} -3 H,
\end{equation}
where we will immediately have
\begin{equation}\label{contin1}
\dot \rho + 3 H (\rho + p) -  \frac {\dot N}{N} (\rho+p) = 0,
\end{equation}
where this equation is introduced as the modified continuity equation by particle creation. In that case, the third sentence provides negative pressure for matter inside the universe which causes the present accelerated expansion. Then, we clearly obtain the solution of Eq. \eqref{contin1} as below
\begin{equation}\label{conteq3}
\rho = C \left(\frac{N}{a^3}\right)^{1 + \omega_m} = C \, n^{1 + \omega_m}.
\end{equation}
where $C$ is an integral constant.

Now we can consider that the matter component within the universe is dominated by the dynamical number of particles $N$, so the matter creation is related to theories of gravity and cosmology. It is interesting to note that researchers consider the negative pressure of Eq. \eqref{contin1} by a viscous fluid, but in this job, we take this issue by the interacting model. The existence of interaction between dark energy and dark matter affects the formation of the structure and evolution of the universe. Therefore, the coupling between dark energy and dark matter leads to the evolution of a uniform distribution of baryons in the formation of the structure of the universe and instability in the dark parts from the early era to the dominance of dark energy in the late era \cite{Zimdahl-2001, Dutta-2018}. This means that the particle creation is considered as an interaction term between the universe components, this is, the particle creation is respectively expressed by the conditions $\mathcal{Q} > 0$ and $\mathcal{Q} < 0$ the source and the sink of energy flow. In that case, by applying $\mathcal{Q} > 0$ and comparing Eqs. \eqref{continuity2-1} and \eqref{contin1}, we find the interaction term $\mathcal{Q}$ with the particle creation in the following form
\begin{equation}\label{mathQ1}
\mathcal{Q} = \frac {\dot N}{N} \rho (1 + \omega_m),
\end{equation}
where the rate of matter production is immediately obtained as
\begin{equation}\label{ratemc1}
\frac {\dot N}{N} = \frac{3 b^2}{1 + \omega_m} H,
\end{equation}
where $N$ is given as follows:
\begin{equation}\label{N2a1}
N = c \, a^{\frac{3 b^2}{1 + \omega_m}}~~~or~~~a = N^{\frac{1 + \omega_m}{3 b^2}},
\end{equation}
where $c$ is an integral constant and we assume it is equal to 1. By substituting Eq. \eqref{N2a1} into Eq. \eqref{conteq3} we obtain the energy density in terms of the scale factor as follows:
\begin{equation}\label{conteq4}
\rho = C \, a^{3 \left(b^2 - 1 - \omega_m \right)},
\end{equation}
where this relation shows us that by applying $b = 0$ and $\omega_m = 0$ we reach the dominated universe with the dust-like matter. Note that Eq. \eqref{conteq4} satisfies Eq. \eqref{continuity2-1}. However, Eq. \eqref{conteq4} is directly derived from \eqref{continuity2-1}, but with the particle creation approach, we can obtain the number of particles, $N$, as an explicit function in the form of Eq. \eqref{N2a1}. On the other hand, we write down the density parameter, $\Omega_m$, as follows:
 \begin{equation}\label{Omegam1}
\Omega_m = \frac{\rho}{\rho_{c}},
\end{equation}
where $\rho_c = 3 H^2 / \kappa^2$ is the critical density. And we have for the present density parameter as
\begin{equation}\label{Omegam2}
\Omega_{m_0} = \frac{\kappa^2 \rho_0}{3 H_0^2},
\end{equation}
where $H_0 = 67.4 \pm 0.5 \, km\,s^{-1}\,Mpc^{-1}$ is the current Hubble parameter, and here we take $\Omega_{m_0} = 0.315 \pm 0.007$ \cite{Aghanim-2017}. The matter energy density, $\rho$, is given by Eqs. \eqref{conteq4} and \eqref{Omegam2} as
\begin{equation}\label{conteq5}
\rho = \frac{3 H_0^2 \Omega_{m_0}}{\kappa^2} \left(\frac{a}{a_0}\right)^{3 \left(b^2 - 1 - \omega_m \right)},
\end{equation}
where $a_0 = 1$ is the current scale factor.

\section{Observational constraints with Hubble parameter data}\label{V}

As mentioned in the previous section, we considered an open system from the thermodynamic point of view for an adiabatic FRW universe model, where there is no heat exchange with the outside universe but the number of particles within the system is created. Therefore, the presence of particles within the universe leads to a modified continuity equation that is related to the interaction term compared to the continuity equation of $f(T)$ gravity. With this perspective, we earned the number of particles $N$ as a function of the scale factor in the form of Eq. \eqref{N2a1}. Therefore, in this work, the perspective or the particle creation from the point of view of $f(T)$ gravity is considered.

In what follows, we insert Eq. \eqref{conteq4} into Eqs. \eqref{fried3} and then we'll have
\begin{subequations}\label{fried4}
\begin{eqnarray}
& \kappa^2 \rho_{de} = 6 H^2 \partial_T f + \frac{1}{2} f - 3 H_0^2\, \Omega_{m_0} \, a^{3\left(b^2 - 1 - \omega_m \right)},\label{fried4-1}\\
& \kappa^2 {p}_{de} = -2 \left(\dot{H} + 3 H^2 \right) \partial_T f + 24 H^2 \dot{H} \partial_{T T} f - \frac{1}{2} f - 3 \omega_m\, H_0^2\, \Omega_{m_0} \, a^{3 \left(b^2 - 1 - \omega_m \right)} - 3 \kappa^2 \xi H. \label{fried4-2}
\end{eqnarray}
\end{subequations}

In order to solve the aforesaid Friedmann equations, we need to find the form of function $f(T)$ in terms of torsion scalar. For this purpose, if we compare Friedmann's Eq. \eqref{fried1-1} with the standard form of Friedman' equation, we will find function $f(T)$ as $f(T) = T - \alpha \sqrt{T}$. Since we are dealing with the modified Friedman equation \eqref{fried1-1}, the form of function $f(T)$ is expected to be different from the form of function $f(T) = T - \alpha \sqrt{T}$. For this purpose, we take the form of function $f(T)$ as follows:
\begin{equation}\label{fT1}
f(T) = T + \alpha T \left(1 - e^{\frac{\beta T_0}{T}} \right) - \alpha T_0 \sqrt{\frac{T}{\beta T_0}} \, \ln\left(\frac{\beta T_0}{T}\right),
\end{equation}
where $\alpha$ and $\beta$ are dimensionless parameters, and $T_0 = -6 H_0^2$ is the current amount of torsion scalar which arises from Eq. \eqref{eamu1}. The above particular model is expected to demonstrate the crossing of the phantom divided-line, which can solve dark energy as a serious challenge in cosmology. It should be noted that this choice is mentioned in Refs. \cite{Bamba-2011, Cai-2016}.

In order to solve the current system, we substitute Eqs. \eqref{eamu1}, \eqref{N2a1}, and \eqref{fT1} into Eqs. \eqref{fried4} and then we will have
\begin{subequations}\label{fried5}
\begin{eqnarray}
& \rho_{de} = \frac{3 \alpha}{\kappa^2}\left(2 \beta H_0^2 - \delta^2 \frac{\dot{N}^2}{N^2}\right) e^{\sigma \frac{N^2}{\dot{N}^2}} + \frac{3 \delta^2}{\kappa^2}\frac{\dot{N}^2}{N^2} \left(\alpha + 1 + \frac{2 \alpha \beta \sqrt{\beta} H_0}{\delta}\frac{N}{\dot{N}}\right) - \frac{3 H_0^2 \Omega_{m_0}}{\kappa^2} N^{3 \delta \eta},\label{fried5-1}\\
& {p}_{de} = \frac{\alpha}{\kappa^2}\left[\delta (3 \delta - 2) \frac{\dot{N}^2}{N^2} - 2 \beta H_0^2 \left(3 - \frac{1}{\delta}\right) + 2 \delta \frac{\ddot{N}}{N} -  \frac{4 \beta^2 H_0^4}{\delta^3} \frac{N^2}{\dot{N}^2} \left(1 - \frac{N \ddot{N}}{\dot{N}^2}\right) - \frac{2 \beta H_0^2}{\delta} \frac{N \ddot{N}}{\dot{N}^2}\right] e^{\sigma \frac{N^2}{\dot{N}^2}} \notag \\
& - \frac{\delta (\alpha + 1)(3 \delta - 2)}{\kappa^2}\frac{\dot{N}^2}{N^2} - 3 \xi \delta \frac{\dot{N}}{N} - \frac{2 \delta (\alpha + 1)}{\kappa^2}\frac{\ddot{N}}{N} - \frac{3 H_0^2 \omega_m \Omega_{m_0}}{\kappa^2} N^{3 \delta \eta} - \frac{2 \alpha H_0}{\kappa^2 \sqrt{\beta}} \left((3 \delta - 1) \frac{\dot{N}}{N} + \frac{\ddot{N}}{\dot{N}}\right), \label{fried5-2}
\end{eqnarray}
\end{subequations}
where the dot symbol demonstrates the derivative with respect to time evolution, and we have $\eta = b^{2} - 1 - \omega_{m}$, $\delta = (1 + \omega_m)/3 b^2$, and $\sigma = \beta H_0^2/\delta^2$. Here we can see that the Friedmann's equations depend on the number of particles, $N$, and its corresponding derivatives with respect to time evolution.

\begin{figure}[h]
\begin{center}
{\includegraphics[scale=.35]{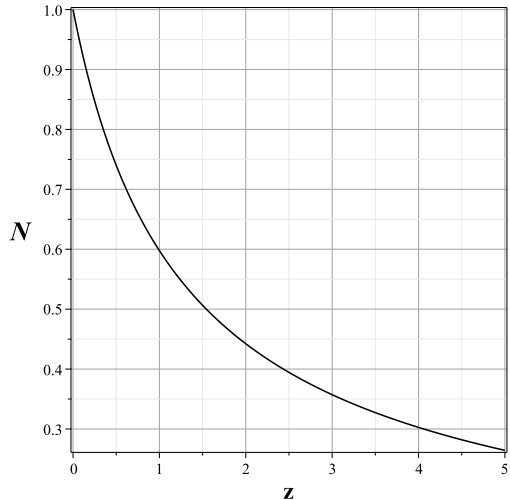}}
\caption{The number of particles in terms of redshift parameter for $\alpha = 0.001$, $\beta = 0.5$, $b = 0.5$, and $\omega_m = 0.01$.}\label{Fig1}
\end{center}
\end{figure}

Now we intend to reconstruct the aforesaid cosmological quantities in terms of the redshift parameter, $z$, which is written in terms of the scale factor by the relation $a(t) = 1/(1+z)$. We can write down the transformation between the derivative with respect to $z$ and the derivative with respect to time evolution as $d/dt = -H (1+z) d/dz$. Therefore, Eq. \eqref{N2a1} is in terms of the redshift parameter as
\begin{equation}\label{Nz1}
N = \left(1+z\right)^{-\frac{1}{\delta}},
\end{equation}
also, we write $\dot{N}$ and $\ddot{N}$ in terms of the redshift parameter in the following form
\begin{subequations}\label{convert1}
\begin{eqnarray}
& \dot{N}(t) = - H (1+z) N', \label{convert1-1}\\
& \ddot{N}(t) = H H' (1+z)^2 N' + H^2 (1+z) N' + H^2 (1+z)^2 N''(z), \label{convert1-2}
\end{eqnarray}
\end{subequations}
where the prime symbol indicates the derivative with respect to the redshift parameter. We can see the variation of the number of particles in terms of the redshift parameter as shown in Fig \ref{Fig1}. Fig. \ref{Fig1} shows us that the value of the number of particles increases from the early universe to the late universe.

In what follows, in order to solve the above scenario, we take a particular model so-called the power-law cosmology as below
\begin{equation}\label{at1}
a(t) = \left(\frac{t}{t_0}\right)^m,
\end{equation}
where $t_0$ is the present age of the universe, and $m$ is a dimensionless positive coefficient \cite{Shafer-2015, Tutusaus-2016}. It should be noted that if $t =t_0$ we will have $a(t_0) = a_0 =1$ which has already been mentioned. The Hubble parameter yields
\begin{equation}\label{hubpar1}
H = \frac{m}{t},
\end{equation}
where the present age of the universe in terms of the present Hubble parameter, $H_0$, is written as
\begin{equation}\label{hubpar2}
t_0 = \frac{m}{H_0},
\end{equation}
where coefficient $m$ is introduced as a correction factor. In that case, the Hubble parameter obtains in terms of the redshift parameter in the following form
\begin{equation}\label{hubpar4}
H = H_0 (1+z)^{\frac{1}{m}},
\end{equation}
where includes only one free parameter, which in turn is less complicated and can be easily calculated with observational constraints.

Since Eq. \eqref{hubpar4} acquired by the power-law model in terms of the redshift parameter, then we have to deliberate the correctness of using it in our model. For this purpose, we fit the power-law model with 51 supernova data from observational Hubble parameter data \cite{Blake_2012, Font_2014, Delubac_2015, Alam_2016, Moresco_2016, Farooq_2017, Pacif_2017, Magana_2018}, then by using the likelihood analysis (the chi-square value, $\chi_{min}^2$), we obtain the value of $m = 0.95$ for these data. In that case, by using Eq. \eqref{hubpar2}, the age of the universe obtains as
$t_0 = 13.78 \,Gyr$ (note that the results of this paragraph are exactly in the Ref. \cite{Mahichi-2021} and we will not go into details in writing).

Now we can obtain $\dot{N}$ and $\ddot{N}$ in terms of the pure redshift parameter in the form
\begin{subequations}\label{convert2}
\begin{eqnarray}
& \dot{N}(t) = \frac{H_0}{\delta} (1+z)^{\frac{1}{m}-\frac{1}{\delta}}, \label{convert2-1}\\
& \ddot{N}(t) = \frac{H_0^2 (m - \delta)}{m \,\delta^2}(1+z)^{\frac{2}{m}-\frac{1}{\delta}}. \label{convert2-2}
\end{eqnarray}
\end{subequations}

By inserting Eqs. \eqref{convert2} into Eqs. \eqref{fried5} we have
\begin{subequations}\label{fried6}
\begin{eqnarray}
& \rho_{de} = \frac{3 \alpha H_0^2}{\kappa^2}\left(2 \beta - (1+z)^{\frac{2}{m}}\right) e^{\beta (1+z)^{-\frac{2}{m}}} + \frac{3 (\alpha+1) H_0^2}{\kappa^2} (1+z)^{\frac{2}{m}} + \frac{6 \alpha \beta \sqrt{\beta} H_0^2}{\kappa^2} (1+z)^{\frac{1}{m}}\notag \\
& - \frac{3 H_0^2 \Omega_{m_0}}{\kappa^2} (1+z)^{-3 \eta},\label{fried6-1}\\
& {p}_{de} = \frac{\alpha H_0^2}{\kappa^2 m} \left(-4 \beta^2 \left(1+z \right)^{-\frac{2}{m}} + (3 m - 2) \left(1+z \right)^{\frac{2}{m}} - 2 \beta (3 m - 1)\right) e^{\beta  \left(1+z \right)^{-\frac{2}{m}}} \notag \\
& - \frac{3 H_0^2 \omega_m \Omega_{m_0}}{\kappa^2} \left(1+z \right)^{-3 \eta} - \frac{H_0}{\kappa^2 m \sqrt{\beta}} \left(\kappa^2 m \sqrt{\beta}\, \xi  + 2 \alpha H_0 \left(3m - 1\right)\right) \left(1+z \right)^{\frac{1}{m}} \notag \\
& - \frac{\left(3 m - 2\right) \left(\alpha +1\right) H_0^2}{\kappa^{2}m } \left(1+z \right)^{\frac{2}{m}}, \label{fried6-2}
\end{eqnarray}
\end{subequations}
where shows that $\rho_{de}$ and $p_{de}$ are explicit functions of the redshift parameter.  We note that these obtained relationships, i.e., the energy density and the pressure of dark energy depend on a few free parameters which arise from scenarios of $f(T)$ gravity, the matter, the particle number, and viscous fluid.

\begin{figure}[h]
\begin{center}
{\includegraphics[scale=.35]{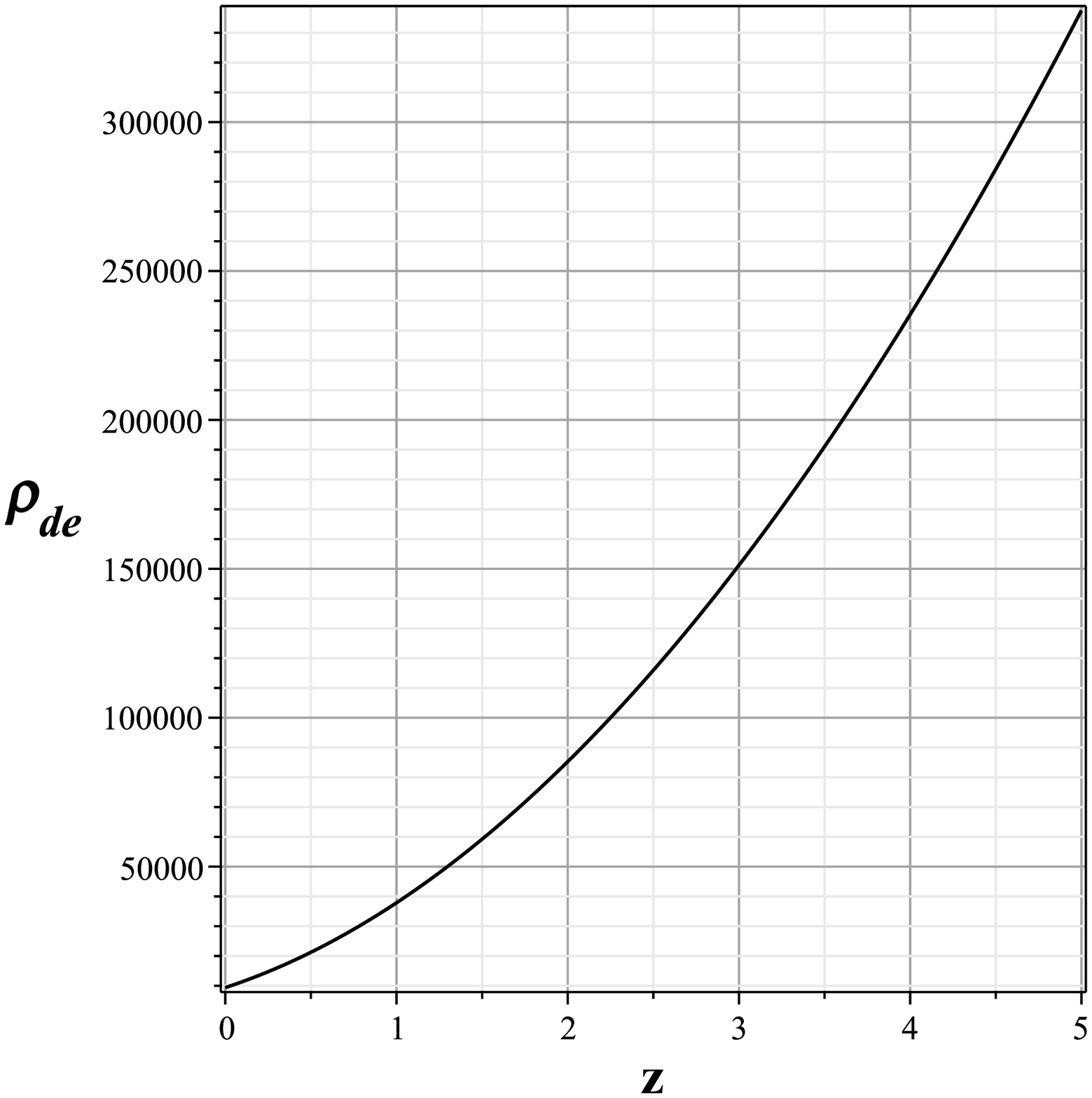}}
{\includegraphics[scale=.35]{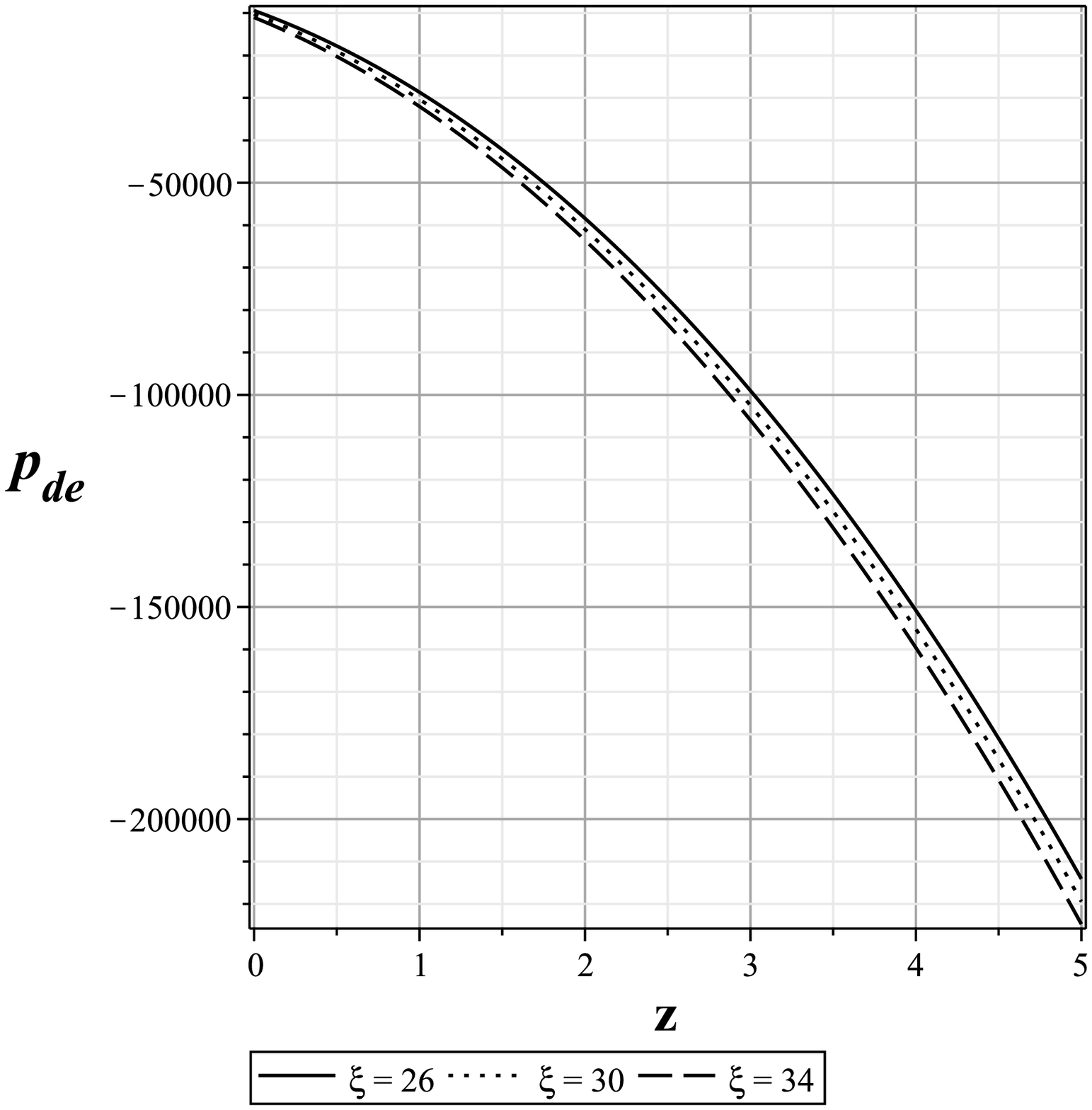}}
\caption{The energy density and the pressure of dark energy in terms of redshift parameter for $\alpha = 0.001$, $\beta = 0.5$, $b = 0.5$, $\omega_m = 0.01$, $\xi = 26$ (line), $\xi = 30$ (dot), and $\xi = 34$ (dash).}\label{rhop1}
\end{center}
\end{figure}

The variation of relationships $\rho_{de}$ and $p_{de}$ are plotted in terms of the redshift parameter as shown in Figs. \ref{rhop1}. It should be noted that free parameters play an important role in drawing the corresponding graphs. For this purpose, we try to choose them by applying conditions such as $\rho_{de} > 0$ and $p_{de} < 0$ especially in the late time ($z = 0$). However, we consider the amounts of free parameters for the corresponding graphs as $\alpha = 0.001$, $\beta = 0.5$, $b = 0.5$, $\omega_m = 0.01$, and $\xi = 26, 30, 34$. Fig. \ref{rhop1} shows us that the value of the energy density decreases from an amount of very high positive in the early time to an amount of much lower positive in the late time ($z = 0$), also the value of pressure decreases from an amount of very large negative in the early time to an amount of much lower negative in the late time ($z = 0$).

\begin{figure}[t]
\begin{center}
{\includegraphics[scale=.35]{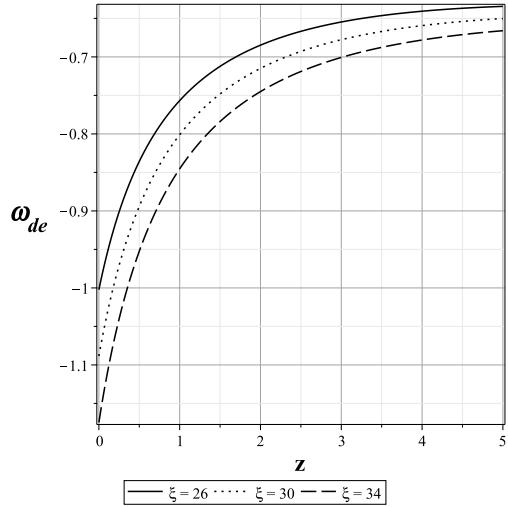}}
\caption{The EoS of dark energy in terms of redshift parameter for $\alpha = 0.001$, $\beta = 0.5$, $b = 0.5$, $\omega_m = 0.01$, $\xi = 26$ (line), $\xi = 30$ (dot), and $\xi = 34$ (dash).}\label{omegade1}
\end{center}
\end{figure}

Since the dark energy EoS is one of the most practical cosmic parameters, it can describe the nature of dark energy and dark matter for different eras of the universe from the Big Bang to the late time. Now using $\rho_{de}$ and $p_{de}$, we clearly find the dark energy EoS as $\omega_{de} = p_{de}/\rho_{de}$ by Eqs. \eqref{fried6} (we avoid writing the corresponding EoS because it is long). In that case, we can plot the variation of the dark energy EoS, $\omega_{de}$, in terms of the redshift parameter as shown in Fig. \ref{omegade1}. Fig. \ref{omegade1} displays us that the amount of EoS decreases from early time to late time, i.e., the evolution of the universe enters from the matter era ($\omega_{de} > -1$) to the phantom era or dark energy era ($\omega_{de} < -1$). Interestingly, the amount of EoS depends on the effect of the viscous fluid, so from Fig. \ref{omegade1} we can see that when the amount of viscosity coefficient increases, the dark energy period occurs with more growth. Therefore, this result indicates that the universe is undergoing an accelerated expansion phase, which is compatible with the obtained results in Refs. \cite{Scolnic_2018} for EoS.

\section{Stability analysis}\label{VI}

\begin{figure}[t]
\begin{center}
{\includegraphics[scale=.35]{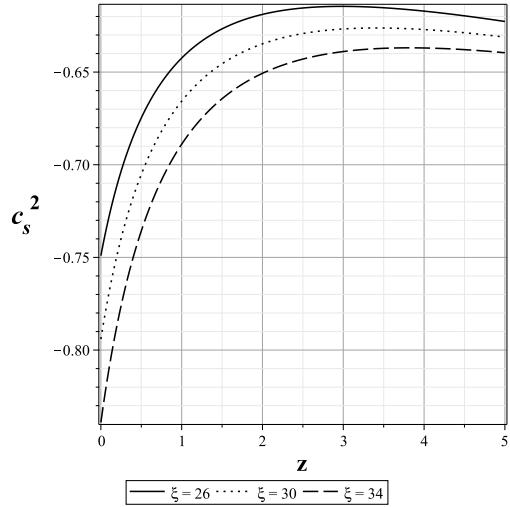}}
\caption{The sound speed parameter in terms of redshift parameter for $\alpha = 0.001$, $\beta = 0.5$, $b = 0.5$, $\omega_m = 0.01$, $\xi = 26$ (line), $\xi = 30$ (dot), and $\xi = 34$ (dash).}\label{cs1}
\end{center}
\end{figure}

In this section, we are going to study the stability and instability of the present model from the thermodynamic perspective. In that case, we use a useful function called the sound speed parameter, $c_s^2$. Since the universe is an adiabatic thermodynamic system, that is, heat energy or mass is not transferred from inside to outside of the universe. This means that the entropy perturbation is zero, and the pressure changes in terms of the energy density. Therefore, we introduce the pressure-to-density ratio with the below sound speed parameter
\begin{equation}\label{cs21}
c_s^2=\frac{\partial p_{de}}{\partial \rho_{de}} = \frac{\partial_z p_{de}}{\partial_z \rho_{de}},
\end{equation}
where symbol $\partial_z$ represents the derivative with respect to the redshift parameter. Now parameter, $c_s^2$, helps us in what range of the redshift parameter is the universe in a state of stability or instability? It should be noted that stability and instability analysis are expressed with conditions $c_s^2 > 0$ and $c_s^2 < 0$, respectively. The variation of the sound speed parameter is plotted in terms of the redshift parameter as shown in Fig. \ref{cs1}. Fig \ref{cs1} shows us that the universe is in a phase of instability at all eras, even with the existence of viscous fluid. This instability indicates that the energy density of dark energy is not in a controllable growth.

\begin{figure}[t]
\begin{center}
{\includegraphics[scale=.35]{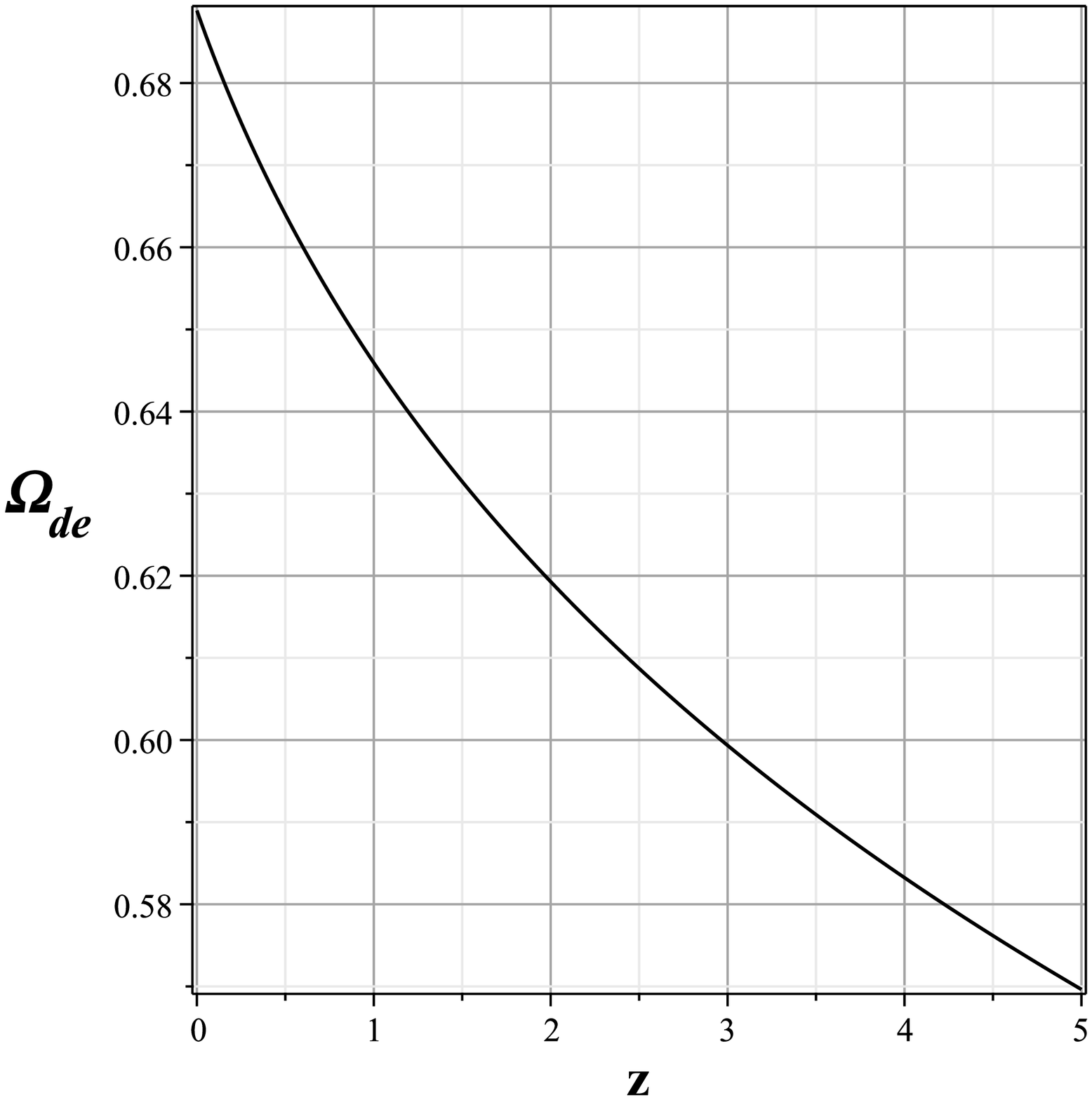}}
{\includegraphics[scale=.35]{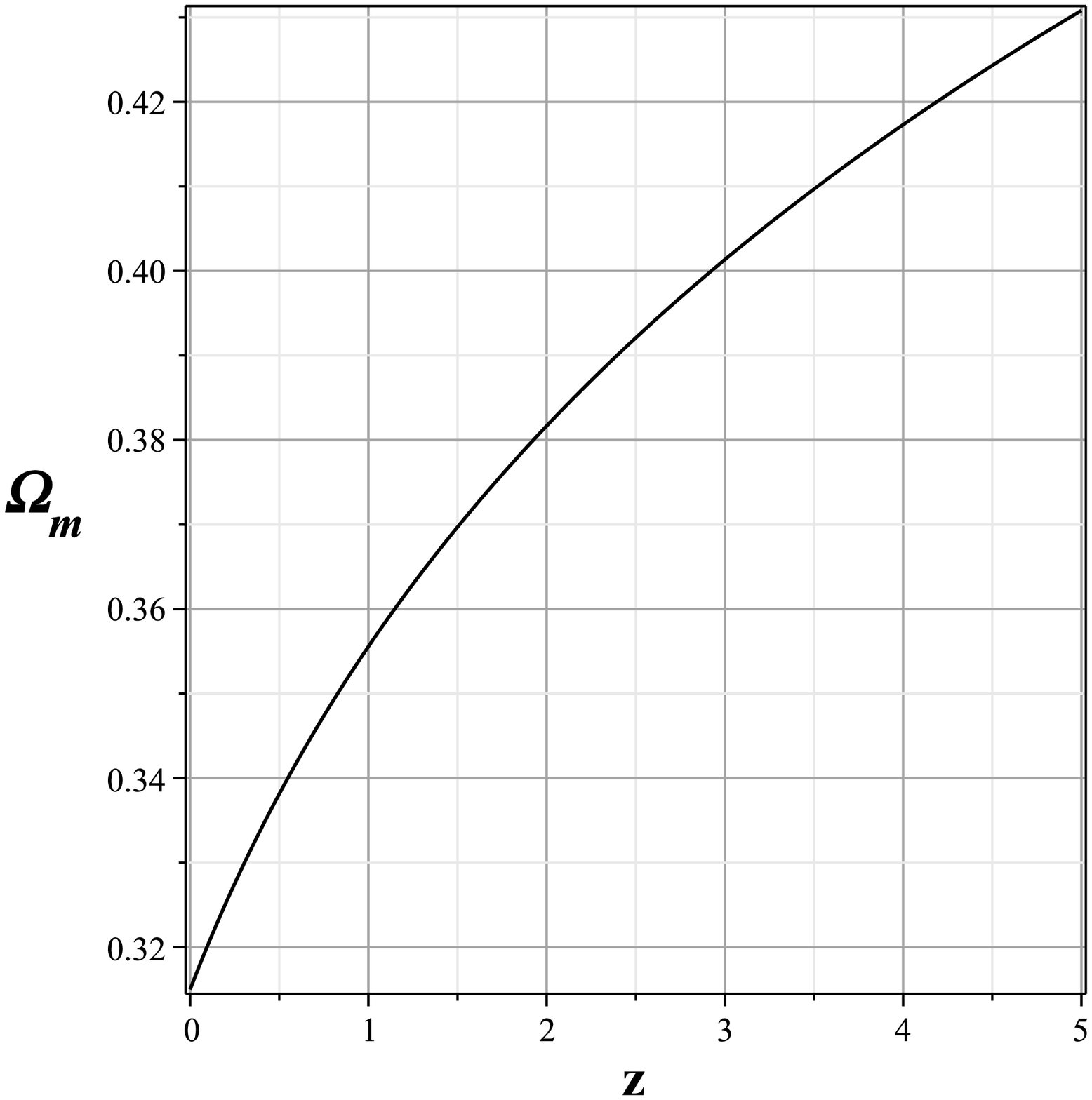}}
\caption{The density parameters of dark energy and dark matter in terms of redshift parameter for $\alpha = 0.001$, $\beta = 0.5$, $b = 0.5$, and $\omega_m = 0.01$.}\label{dedm1}
\end{center}
\end{figure}

Now for a more complete description of our model, we try to calculate the present values of the density parameters for dark energy and dark matter. According to the density parameter for dark matter \eqref{Omegam1}, the density parameter for dark energy is introduced in the form
\begin{equation}\label{Omegade1}
\Omega_{de} = \frac{\rho_{de}}{\rho_c},
\end{equation}
where $\rho_c$ is the critical density. By inserting Eq. \eqref{conteq5} into Eq. \eqref{Omegam1}, the variation of the density parameter for dark matter is plotted in terms of the redshift parameter as shown in Fig. \ref{dedm1} to the right. Similarly, we draw the variation of the density parameter for dark energy in terms of the redshift parameter by substituting Eq. \eqref{fried6-1} into Eq. \eqref{Omegade1} as shown in Fig. \ref{dedm1} to the left.
Fig. \ref{dedm1} shows us that the amount of $\Omega_m$ decreases from the early universe to the late universe, which eventually reaches the current value of $0.315$. On the other hand, the amount of $\Omega_{de}$ increases from the early time to the late time, which reaches the present amount of $0.689$. Therefore, these calculations show us that the share of dark energy is increasing, which means that the universe is under a phase of accelerated expansion.


\section{Conclusion}\label{VII}

In this paper, we explored particle creation from the perspective of $f(T)$ gravity by flat-FRW metric in the universe containing viscous fluid. First of all, for obtaining the Einstein equation in $f(T)$ gravity, we used the Weitzenb\"ock connection in teleparallel gravity instead of the Levi-Civita connection in general relativity. The energy-momentum tensor has been considered in terms of the total energy density, the total pressure, and the pressure of bulk viscosity of fluid within the universe. Next, we acquired the Friedmann equations and the EoS of dark energy in terms of a function of torsion scalar $T$, and its first and second derivatives. And then, the continuity equations for the universe components were obtained by considering the interaction model. The interaction term $\mathcal{Q}$ was considered as $\mathcal{Q} = 3 b^2 H \rho$ in which $b^2$ is the transfer strength from dark matter to dark energy.

Another significant point of this paper is that the universe should be considered as an adiabatic system from the perspective of the first law of thermodynamics. For this purpose,  we considered the creation of the number of particles in this universe, so that we obtained the modified continuity equation for particle creation. The interesting point is that the obtained continuity equation when matter creation has a negative pressure which is causing the present accelerated expansion. Next, we acquired the energy density of matter in terms of the number of particles, $N$ as shown in Eq. \eqref{conteq3}. The interesting point of this paper is that Eq. \eqref{contin1} is equivalent to Eq. \eqref{continuity2-1}, This means that we consider the particle creation to be equivalent to the exchange of energy between the components of the universe (same as interaction term $\mathcal{Q})$. The result of this equivalency leads to obtaining the number of particles, and the energy density of matter in terms of the scale factor.

In what follows, we relate the obtained energy density of matter to Friedmann equations of $f(T)$ gravity with a specified function $f(T)$ of Eq. \eqref{fT1}, which is exponential and logarithmic. In this case, the result was to get the energy density \eqref{fried5-1} and the pressure \eqref{fried5-2} of dark energy in terms of the number of particles $N$. Then, we reconstructed the number of particles in terms of the redshift parameter, the benefit of this reconstruction is that the cosmological parameters can be evaluated in the late time ($z = 0$). In order to evaluate the present paper, we took the power-law for the scale factor, so that we immediately acquired the Hubble parameter \eqref{hubpar4} in terms of the redshift parameter. After that, we used the result of Ref. \cite{Mahichi-2021} for fitting the Hubble parameter by the minimum chi-square value with $51$ observational Hubble parameter data. And then, we plotted the graphs of the cosmological parameters such as the energy density, the pressure, and the EoS of dark energy in terms of the redshift parameter. These graphs indicated that the universe is in a phase of accelerated expansion. Then, stability analysis of the present job was investigated by the sound speed that the corresponding graph showed the universe is in an instability phase for the late time, i.e., the growth of the energy density of dark energy is not controlled in the present universe. Finally, in order for the results to be consistent with the observational data, we calculated the density parameter values for dark energy and dark matter with values of $0.689$ and $0.315$, respectively, and also plotted them against the redshift parameter.


\end{document}